\documentclass[sigconf]{acmart}

\usepackage{caption}
\usepackage{subcaption}
\usepackage{amsmath}
\usepackage{array}
\usepackage{graphicx}
\usepackage{amsthm}
\usepackage{booktabs}
\usepackage{float}
\usepackage{mathtools}
\usepackage{xcolor}
\usepackage{bm}
\usepackage{hyperref}
\usepackage{enumitem}
\setlist{nolistsep}
\usepackage{listings}
\usepackage{multicol}
\usepackage{multirow}
\usepackage{tikz}
\usetikzlibrary{shapes,arrows,arrows.meta,positioning}
\tikzset{%
  >={Latex[width=2mm,length=2mm]},
            base/.style = {rectangle, rounded corners, draw=black,
                           minimum width=3cm, minimum height=1cm,
                           text centered, font=\sffamily},
  activityStarts/.style = {base, fill=blue!30, text width=3cm},
       startstop/.style = {base, fill=red!30, text width=3cm},
    activityRuns/.style = {base, fill=green!30, text width=3cm},
         process/.style = {base, minimum width=2.5cm, fill=orange!15,
                           font=\ttfamily, text centered, text width=3cm},
}

\begin{document}

\title{Predicting the Effect of European Air Traffic on Cirrus Cloud Cover}

\author{T. van der Duim, M. Chekol}

\begin{abstract}
    The purpose of this study is to provide more insight into the role of anthropogenic cirrus formation through air traffic, by investigating the high-density European airspace over a period spanning several recent years including the start of the COVID-19 pandemic (2015-2020). Several data resources are combined, exploiting the strengths of each product within an all-encompassing framework on a high spatio-temporal resolution. Data from METEOSAT SEVIRI have been combined and validated with CALIPSO's CALIOP data to deduce temporal cirrus cloud cover variability over a rectangular region bound by ($10^{\circ}$W-$35^{\circ}$N) and ($40^{\circ}$E-$60^{\circ}$N). Cirrus clouds are correlated with air traffic. Meteorology was incorporated into the analysis as it is of major influence on the formation and lifetime of cirrus. Both a logistic regression model and a Random Forest model were built to assess cirrus cloud cover variability imposed by meteorology. The impact of aviation on cirrus cover in 1) supersaturated an 2) sub-saturated air have been evaluated separately. A description of all the datasets involved, including the main research methodology and main results, are presented.
\end{abstract}

\maketitle

\section{Introduction}

Cirrus clouds (\textit{CC}) are ubiquitous clouds composed of ice crystals that are long-lived under the right conditions, and that reside predominantly in the upper troposphere and lower stratosphere \citep{tropostrato}. Their major importance relates to their effect on Earth's radiative balance, being the balance between incoming solar radiation and outgoing planetary radiation. High, optically thin (highly transparent) clouds like cirrus have the tendency to be nearly transparent to incoming shortwave radiation, while they are opaque to planetary longwave radiation \citep{radiation}. Hence they tend to trap and reflect down more radiation than is reflected or re-emitted into space, leading to a net warming effect on the lower atmosphere.

\textit{CC} can be formed either naturally or due to anthropogenic activity, in the latter case due to air traffic (aircraft-induced cirrus or \textit{AIC}). Out of aviation-derived radiative forcing components, the \textit{AIC} component is estimated to be potentially larger than the direct component from greenhouse gases, particularly CO$_2$ and NO$_x$ \citep{formation}. These observations, in combination with low confidence estimates of cloud radiative forcing feedbacks, exemplify the need of a thorough assessment of cirrus cloud properties (\textit{CCP}) and the way those properties are affected by anthropogenic activity.

\begin{figure}[t]
    \centering
    \includegraphics[width=0.5\textwidth]{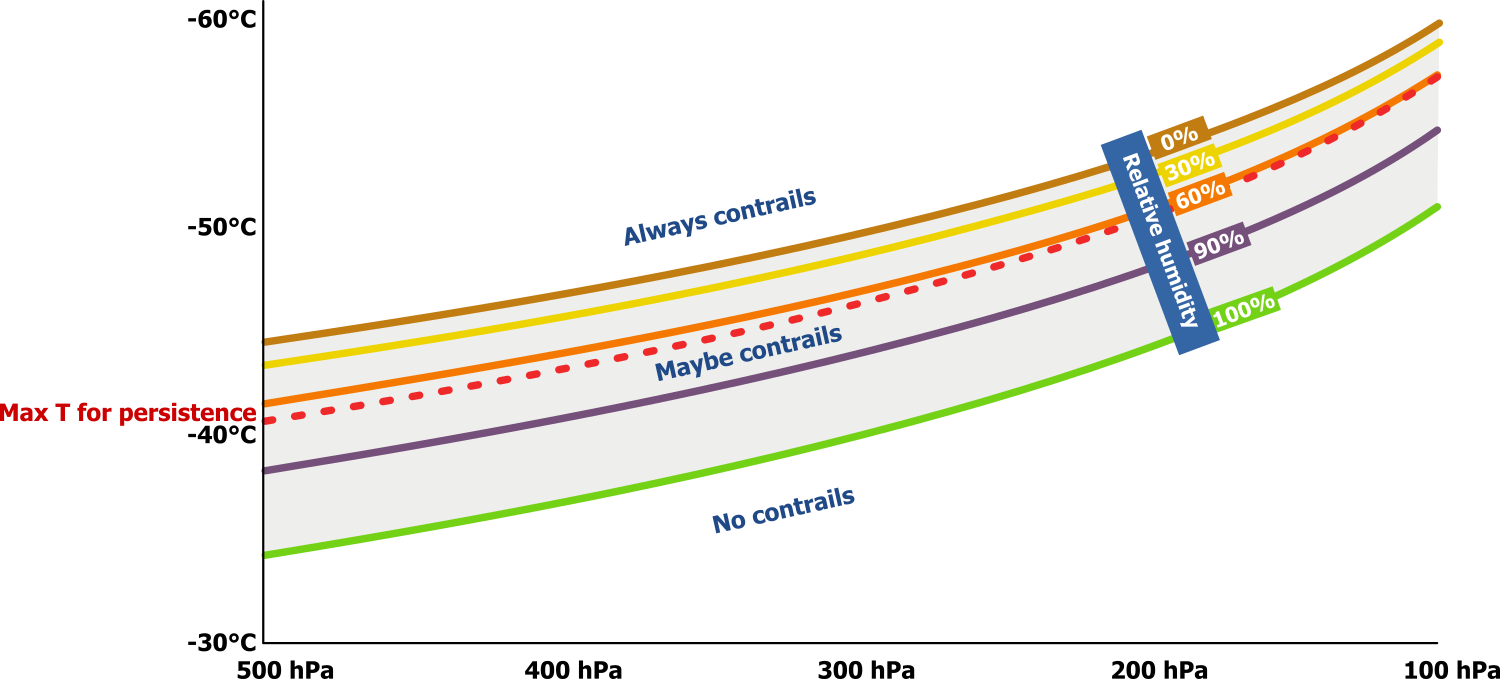}
    \caption{Appleman chart. Air pressure level is shown on the horizontal axis, ambient temperature on the vertical axis. Retrieved from \citep{appleman}.}
    \label{fig:appleman}
\end{figure}

The meteorological state of the atmosphere plays a key role in the formation and on the lifetime of \textit{CC}. \textit{CC} can only be formed and prevail if the air is supersaturated with respect to ice \citep{formation}. Ice supersaturation refers to the condition where, in this case the air, exceeds the water vapor threshold that would be needed to produce saturation relative to pure ice, that is, the relative humidity (\textit{RH}) is greater than 100\%. Ice supersaturation can occur under sufficiently cold and moist conditions, and its degree together with the presence of atmospheric ice nuclei (particles which act as nuclei for ice crystals to form and grow) dictate the chance of \textit{CC} occurrence. The Schmidt-Appleman criterion, which formulates the conditions under which supersaturation takes place, can be used to assess whether atmospheric conditions are sufficiently cold and humid for \textit{AIC} to form, see Figure~\ref{fig:appleman}.

\begin{figure}[t]
    \centering
    \includegraphics[width=0.4\textwidth]{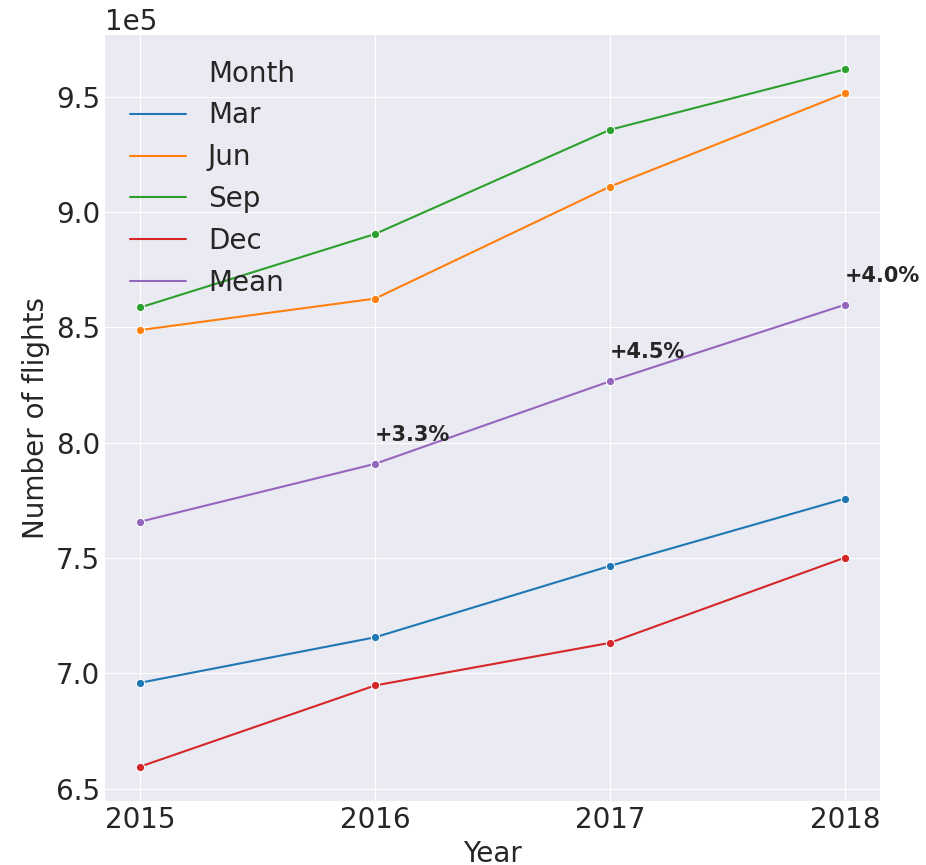}
    \caption{Total number of flights per month detected by European radar, for four months a year over the period 2015-2018. Data derived and analyzed from EUROCONTROL OneSky database [retrieved on Jan 12 2021].}
    \label{fig:air_traffic_trend}
\end{figure}

This research builds on state-of-the-art satellite products, a big data approach that uses machine learning, on high resolution data to retrieve more insights into the relationship between air traffic and cirrus cloud cover (\textit{CCC}). The focus is on European airspace due to its high air traffic density and data coverage, bounded by ($10^{\circ}W$, $35^{\circ}N$) and ($40^{\circ}E$, $60^{\circ}N$). Temporal patterns in European air traffic, including the persistent rise in global European air traffic \citep{passengers} (Figure~\ref{fig:air_traffic_trend}) up till the COVID pandemic and subsequent pandemic, with exceptional low air traffic density, motivate to study time series analysis of \textit{CCC} spanning multiple recent years.

Besides meteorology and air traffic density, engine combustion type also relates to cirrus formation \citep{contrailtemp}. The engine type determines engine combustion properties including some of the key parameters in cirrus formation, being exhaust temperature, water vapor content and the expulsion of particles that could function as ice nuclei on which air particles could crystallize. For example, \cite{contrailtemp} argues that high by-pass engines lower the exhaust temperature leading to sooner (i.e. lower temperature threshold) contrail formation.
In summary, within this study, we answer the following research questions:
\begin{enumerate}
    \item How does contrail formation evolve over time between 2015 and 2020 under 1) increasing air traffic and 2) a global pandemic?
    \item How large is the impact of air traffic on \textit{CCC} and does combustion type significantly alter the results?
\end{enumerate}
Proper assessment of the stated research questions require integration of various data sources, as well as accurate handling of meteorology which, based on literature, has a strong interplay with \textit{CCC}.

\section{Related work}

The study on contrail formation goes back to the year 1919 \citep{contrailtemp}. Nonetheless, current day research on contrail formation finds its roots in the post-World War II period in which \citep{condens} published an article on the formation of condensation trails by jet aircraft. The theoretical foundation laid down in this research relates to what is frequently referred to in current literature as the Schmidt-Appleman theory \citep{propeff, aircraftinduced, ocean}. Within this theory it was well-established that for \textit{CC} to form and prevail, the ambient vapor pressure should be greater than the saturation vapor pressure with respect to ice, as long as temperatures are below the corresponding critical temperature \citep{formation}. Scientific consensus has ever since been reached about the framework that \textit{CCC} is a function of air traffic exhaust properties, air pressure, humidity and ambient temperature.

A similar study to the one presented here \citep{aircraftinduced}, focusing on the attribution of air traffic to \textit{CCC} in European airspace and using EUROCONTROL air traffic data together with METEOSAT data on \textit{CCC}, reported a significant relationship between air traffic and \textit{CCC}. Despite the similar parametrization of air traffic density compared to the presented study in this paper, the spatial resolution on which the parameter is calculated is much lower in the previous work ($1^{\circ}\times 1^{\circ}$ vs $0.25^{\circ}\times 0.25^{\circ}$). Also, engine exhaust properties and the limited detectability of the used satellite product regarding optically thin \textit{CC} (optical depths below 0.2) are disregarded in the previous study. Most studies up till now relied on space-born LIDAR data from the CALIPSO satellite, as LIDAR gives a means to accurately detect \textit{CC} from space, including optically thin cirrus.

Previous studies consistently reported a significant impact of air traffic on \textit{CCC} and \textit{CC} optical depth \citep{alreadyexisting, changescirrus}. Nonetheless, the estimates in (relative) magnitude of the anthropogenic influence on \textit{CC} properties are highly variable. Together with the associated climate impact this stipulates the need of a more thorough assessment of \textit{CC} forced by air traffic.
\section{Datasets}

This study involves multimodal data that are all publicly available. The various sources comprise approximately 200GB of data. Due to the size of the datasets and their public accessibility, the original sources are referenced throughout this section, from which the datasets can be retrieved. The framework for the analysis, consisting of the source codes, links to the datasets, etc are available on Github~\citep{github}.

\subsection{Space-born Cirrus Detection}

Cirrus clouds are, due to their optically thin nature, not easily detectable by satellites. Over the past decades more satellites have been equipped with sensors that are able to detect cirrus, using the Shortwave-Infrared (\textit{SWIR}) portion of the electromagnetic spectrum at 1.38$\mu$m. At this wavelength a large portion of the radiation is reflected by ice crystals contained in cirrus clouds, while nearly all radiation beneath the cloud is absorbed by water vapor \citep{channel}. The inability of satellites equipped with passive sensors \citep{remotesensors} like MODIS, Landsat and Sentinel-2 to detect optically thin ($\sigma < 0.2$) cirrus clouds is a major drawback for such products. Active satellites \citep{remotesensors}, like the Cloud-Aerosol Lidar and Infrared Pathfinder Satellite Observation (\textit{CALIPSO}) carrying on board the Cloud-Aerosol Lidar with Orthogonal Polarization (\textit{CALIOP}), have the advantage over passive satellites to detect a broader optical range of clouds while not being restricted to daytime sampling. This section provides an overview of the satellite data products involved in this research.

\subsubsection{CALIPSO Satellite}

The CALIPSO satellite is a nadir-viewing active satellite, equipped with the CALIOP lidar that is polarization-sensitive and emits dual-frequency pulses with a repetition rate of 20 Hz. From the backscatter signal CALIPSO is able to construct high-resolution vertical cloud profiles, including optically thin cirrus. Hence, for an optimal representation of cirrus properties one could argue CALIPSO is the most accurate product. A major drawback of CALIPSO for the purpose of this research is its polar orbit, meaning the region of interest (\textit{ROI}) is intermittently observed by the satellite. Together with the limited swath width (width of the laser scanning beam), the temporal resolution is low considering a fixed location on Earth. For this research the \textit{CAL\_LID\_L2\_01kmCLay-ValStage1-V3} \citep{calipso40} product is used, as it combines a high spatial resolution of $0.1\times 0.1$ km with product maturity. The latter refers to the validated stage of the product.

The CALIPSO LIDAR products used are level 2 products, meaning they come in non-uniform spatio-temporal grid scales. The products are formatted in Hierachical Data Format (HDF4). The datasets provide, among a set of 94 variables, a binary-encoded 16-bit integer containing classification flags, i.e. for each detected layer by the lidar backscatter signal it contains 1) feature type (e.g. cloud), 2) feature sub-type (e.g. cirrus) and 3) quality assessment \citep{calipsodesc}. Upon decoding, only the detected cirrus clouds which have been provided a medium or high confidence level (both feature type "\textit{cloud}" and feature subtype "\textit{cirrus}") are kept for the analysis. Additionally, cloud optical thickness (\textit{COT}), cloud mid-layer pressure and cloud mid-layer temperature are extracted. The cloud mid-layer pressure is used as a cloud locator in the vertical to bin the detected cirrus into predefined layers.

For the \textit{ROI} the average overpass time of CALIPSO is 7 minutes. In order to simplify the analysis without comprising its accuracy too much, the mean timestamp during the overpass is taken as fixed timestamp for that overpass, hence leading to a maximum time offset of 3.5 minutes of each individual observation. Moreover, the data is resampled over the spatial domain into a fixed, uniform longitude-latitude grid of $0.25^{\circ}\times 0.25^{\circ}$. As the aforementioned poor temporal resolution cannot be resolved within acceptable uncertainty, this dataset does not suffice for the analysis in its entirety. A complementary data source is required which provides data on a higher temporal resolution. Geostationary satellite METEOSAT provide the means with the SEVIRI imager.

\subsubsection{METEOSAT Satellite}

EUMETSAT's METEOSAT Second Generation (MSG) satellite carries on board the "Spinning Enhanced Visible and InfraRed Imager" (SEVIRI), which is an imaging instrument in geostationary orbit that continuously observes the \textit{ROI}. Eight out of twelve spectral channels operate in the thermal infrared spectrum and generate data on i.a. \textit{CC}. The strength of this product resides in its high temporal and spatial resolution (15 mins and $5\times 5$ km, respectively). However, as it is a passive sensor it misses optically thin clouds, and the product runs up to 2017. Nevertheless, the fact that the relevant channels for cirrus detection operate in the thermal infrared spectrum, make the use of this product possible during the night. The product contains both \textit{CCC} and (cirrus) optical thickness (\textit{COT}).

The CLAAS-2.1 record \citep{claas} provides data derived from SEVIRI on cloud properties. The CLAAS product is validated and inter-calibrated with data from MODIS Aqua. The spatial and temporal resolution of the product corresponds to the native METEOSAT SEVIRI resolution and features cloud type as well as cloud microphysical properties such as \textit{COT}. Each 15-minute mapping is stored in a separate NetCDF file, and an accompanying auxiliary file allows transformation of this level 2 product into a geospatial lon-lat grid mapping. This gridding is done on the same resolution as the aforementioned CALIPSO product, being $0.25^{\circ}\times 0.25^{\circ}$.

The high temporal resolution of the product over the entire \textit{ROI} makes this product an attractive complement to this research. More specifically, this product is used to assess the second research question upon product validation with CALIPSO. For the time series analysis this product does not suffice due to its temporal coverage.

\subsection{Air Traffic}

The EUROCONTROL OneSky R\&D data archive \citep{R&D} provides high resolution data on all commercial flights operating in and over Europe, supplemented with data originating from air navigation service providers' radar and datalink communication. The temporal domain of this data spans the years 2015-2018, and the months March, June, September and December. Those datasets contain all flights that pass through European airspace, i.e. they are not restricted to flights that depart from or arrive at an European airport. Each monthly data bundle i.a. consists of a \textit{Flights} dataset, which contains all flights that were registered during that month, and a \textit{Flight Points Actual} dataset including all filed or actual flight paths. All datasets are in CSV format.

\subsubsection{Flights Dataset} 
Within the \textit{Flights} dataset each row is a unique flight, hence the total number of rows is the total number of registered flights. The datasets include a unique identifier for each flight \textit{ECTL\_ID}, the ICAO departure airport code \textit{ADEP}, the latitude of the departure airport \textit{ADEP Latitude}, the longitude of the departure airport \textit{ADEP Longitude}, the ICAO destination airport code \textit{ADES}, the latitude of the destination airport \textit{ADES Latitude}, the longitude of the destination airport \textit{ADES Longitude}, the planned arrival time \textit{Filed Arrival Time}, the time an aircraft departs from its parking spot \textit{Actual Off-Block Time}, the \textit{Actual Arrival Time}, the aircraft type \textit{AC Type}, the aircraft operator \textit{AC Operator}, the unique aircraft identifier \textit{AC Registration}, the \textit{ICAO Flight Type} (either S - scheduled or N - Non-scheduled), the market segment of the operation \textit{STATFOR Market Segment}, the requested cruising flight level \textit{Requested FL} and the distance flown in nautical miles \textit{Actual Distance Flown (nm)}.

\subsubsection{Flight Points Actual Datasets}
In the \textit{Flight Points Actual} datasets each row is a reported aircraft location at a given time based on radar tracking. The datasets include a unique identifier for each flight \textit{ECTL\_ID}, a numeric sequence number of the points crossed by the flight \textit{Sequence Number}, the time at which the point was passed \textit{Time Over}, the flight level (altitude) at that point \textit{Flight Level}, the latitude at that moment \textit{Latitude} and the longitude at that moment \textit{Longitude}. For March 2015 the \textit{Flights} dataset consists of 698,715 rows (unfiltered) and the \textit{Flight Points Actual} dataset of 17,549,122 rows (unfiltered), meaning flights were tracked about 25 times on average during a single operation.

In order to extend the temporal coverage of air traffic data to COVID-19 lockdown period (2020), another data source is consulted. Worldwide crowdsourced air traffic data from the same OpenSky network has been made available for 2019 and 2020 on Zenodo \citep{zenodo}. From this source only one CSV file is provided for each month, with features highly similar to the aforementioned \textit{Flight Points Actual} datasets, the main difference being that those crowdsourced datasets merely contain two reported locations per flight.

Conditional filters are applied concerning invalid (NaN) entries. Omitting all flights containing invalid arguments would lead to an undesirable data reduction, particularly for the 2019 and 2020 datasets. E.g. any missing values within the \textit{market} or \textit{AC\_operator} column are acceptable, while in case of any missing value in the \textit{Latitude} or \textit{Longitude} column the row would be omitted. The latter applies to on average 0.02\% of all data rows in the \textit{Flight Points Actual} datasets.

The time gaps between which flights are geolocated are highly variable and range from minutes to more than an hour in some cases, albeit around 90\% of the time gaps are 10 minutes or less. Linear interpolation in three spatial dimensions (longitude, latitude and flight level) is performed based on a 30 second time interval, in order to attain a higher resolution later on in the analysis. An even higher sampling frequency was not attainable due to computational limits. Performing the interpolation on data of March 2015 with an original file size of around 18 million flight points results in a file containing roughly 83 million flight points.

Regarding the datasets retrieved for 2019 and 2020 from \citep{zenodo}, which do not contain more detailed along-flight location tracking and include non-European flights, the datasets from 2015 to 2018 are used in the data wrangling process. A dichotomous approach is used to extract the subset of flights that flew through the \textit{ROI}. The unique set of departure-arrival airport pairs over 2018 are used to filter out the flights that flew over the \textit{ROI} in 2019 and 2020 with high probability. In addition, all flights with a departure or arrival location within the \textit{ROI} are included. Upon removing all duplicates (single flights might be detected by both methods) an estimated flight dataset for the considered month is obtained. This approach fails to detect flights that neither have a reported geolocation within the \textit{ROI} nor a pre-occurring departure-arrival combination from previous years, which might be the case for e.g. new flight routes. Any concern should be raised regarding the unofficial source of the data, some unrealistic entries related to flight levels and the many missing values.

A final data filtering approach relates to flight levels. As cirrus clouds do not generally form below 6km, all data points of aircraft flying below this level could safely be filtered out without compromising the analysis. 6km corresponds to a flight level of 197 (hundreds of feet). Compared to the original dataset, around half of the data is removed after performing all those data wrangling steps, the majority taken up by the flight level filter.

Some valuable features for the sake of the research are contained in both the \textit{Flights} dataset (A/C type, departure and destination airport) and in the \textit{Flight Points} dataset (discretized flight path with reported geolocations). The two datasets have one column in common, being '\textit{ID}' and '\textit{ECTL\_ID}'. Using this column the cleaned and processed datasets can be merged into one. These 'merged' datasets are the ones that are used for the analysis.

\subsection{Auxiliary Data}

\subsubsection{Meteorological Datasets} Meteorology plays a central role within this analysis due to its primary influence on the probability of cirrus formation. ERA5 hourly reanalysis data on pressure levels \citep{ERA5} from the European Centre for Medium-Range Weather Forecasts (ECMWF) is extracted on a horizontal resolution of $0.25^{\circ}\times 0.25^{\circ}$ and at various pressure levels. A selection of pressure levels is made, based on the pressure levels at which \textit{CC} may generally form: 100 hPa, 125 hPa, 150 hPa, 175 hPa, 200 hPa, 225 hPa, 250 hPa, 300 hPa, 350 hPa, 400 hPa and 450 hPa, which is roughly in between 7 and 16 km geopotential height, i.e. including the upper troposphere and lower stratosphere. From this data source the ambient air temperature and relative humidity at each pressure level is extracted. The data is formatted into NetCDF files.

\subsubsection{Aircraft Engine Features Datasets}
The Aircraft Performance Database \citep{APB} maintained by EUROCONTROL is a user interface where aircraft details including performance and technicalities can be found. Aircraft can be found based on ICAO code. Using the ICAO codes reported in the \textit{Flights} datasets, the corresponding aircraft metadata can be scraped off the website using a URL retrieval algorithm. This data source is used to retrieve the number of engines and the default engine type mounted on the sought aircraft type. This data source is used in conjunction with the ICAO Aircraft Emissions Databank \citep{icao}, a dataset in CSV format containing over 800 engine types and their specifications, amongst which engine fuel flow.

\section{Spatio-temporal air traffic data analysis}
\label{sec:atspatio-temp}

The flight datasets are parametrized into an interpretative quantity that can be projected onto a fixed two-dimensional grid of $0.25^{\circ}\times 0.25^{\circ}$. This same grid is used to analyze \textit{CC} within the \textit{ROI}, in order to be able to correlate air traffic with \textit{CC}. The parameter relating to air traffic will be referred to as air traffic density (\textit{ATD}) and carries the unit of distance per km$^2$ per hour, in line with \cite{aircraftinduced}. All flight paths are integrated and subsequently aggregated within each grid cell over a defined time window, using the Euclidean distance metric.

\begin{figure}[t]
    \centering
    \includegraphics[width=0.5\textwidth]{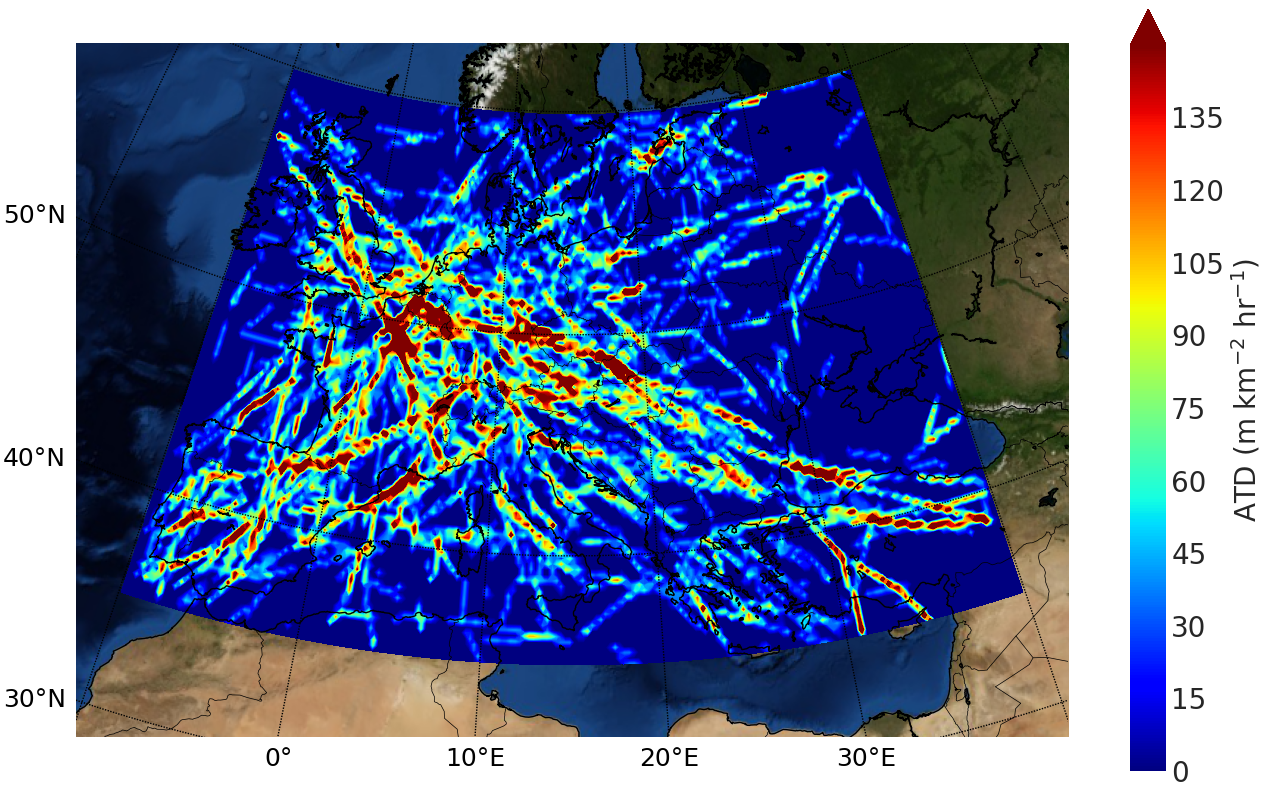}
    \caption{Air Traffic Density shown for the 1st of March 2015 between 12 PM and 12:15 PM. Units are in m km$^{-2}$ hr$^{-1}$. The data has been gridded on a $0.25^{\circ}\times 0.25^{\circ}$ grid.}
    \label{fig:ATD}
\end{figure}

In order to include the vertical spatial dimension, atmospheric boxes are defined as grid cells bounded by two pressure levels (which is a modelling parameter). Another control parameter is the time window over which \textit{ATD} is computed. Increasing this window masks some spatial-temporal variability in \textit{ATD}, albeit its implications depend on the purpose of the analysis. In any case the time window should be larger than the discretization in time of the flight paths, as flights that pass through the box might be completely missed otherwise. This readily exemplifies the importance of interpolating the flight paths as described previously. Figure \ref{fig:ATD} shows a contour map of \textit{ATD} as retrieved by the algorithm on the 1st of March 2015 between 12 PM and 12:15 PM, for all flight points between 6 and 14 km altitude. The data has been interpolated in time with $\Delta t = 30$ seconds, where some major European flight hubs (Paris, Brussels) are clearly detectable. As this data only include flight levels of 197 of higher, not all airport hubs stand out. The linear pathways partially emerge due to the existence of flight corridors and partially due to the linear interpolation method.

\begin{figure*}[ht!]
\begin{tikzpicture}[node distance=1.5cm,
every node/.style={scale=0.7,fill=white, font=\sffamily}, align=center]
\node (start) [process] {extract A/C type};
    \node (euro) [activityStarts, above of=start, yshift=0.5cm] {EUROCONTROL datasets};
    \node (in1) [process, right of=start, xshift=3.8cm] {link A/C type to A/C metadata};
    \node (APB) [activityStarts, above of=in1, yshift=0.5cm] {Aircraft Performance Database webpage};
    \node (in2) [process, right of=in1, xshift=3.8cm] {interpret metadata using NLP algorithms};
    \node (in3) [process, above of=in2, yshift=1cm] {get number of engines};
    \node (in4) [process, below of=in2, yshift=-1cm] {get engine type};
    \node (in6) [process, right of=in4, xshift=3.8cm] {get fuel flow engine};
    \node (in5) [activityStarts, right of=in6, xshift=3cm] {ESA ICAO Aircraft Emissions Databank};
    \node (in7) [startstop, right of=in2, xshift=3.8cm] {fuel flow aircraft (41 or 32\%)};
    
    \draw[->] (start) --node[text width=1cm] {130 (100\%)} (in1);
    \draw [->] (euro) -- (start);
    \draw [->] (in1) --node[text width=1cm] {115 (89\%)} (in2);
    \draw [->] (APB) -- (in1);
    \draw [->] (in2) --node[text width=1cm] {108 (83\%)} (in3);
    \draw [->] (in2) --node[text width=1cm] {108 (83\%)} (in4);
    \draw [->] (in4) --node[text width=1cm] {108 (83\%)} (in6);
    \draw [->] (in5) -- (in6);
    \draw [->] (in6) --node[text width=1cm] {42 (32\%)} (in7);
    \draw [->] (in3) -|node[text width=1cm] {108 (83\%)} (in7);
\end{tikzpicture}
  \caption{Flowchart of retrieval algorithm fuel flow. The numbers at the arrows show the total number of aircraft types contained in the analysis for March 2015. The percentages are taken w.r.t. the total of 130 aircraft types.}
    \label{fig:flowchartff}
\end{figure*}
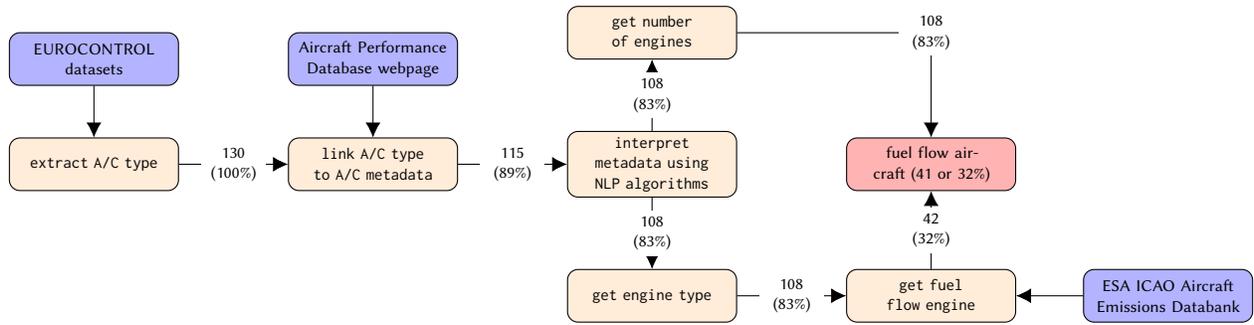

A more accurate approach which would better fit the analysis is to parametrize air traffic by taking into account the engine exhaust properties. The reasoning behind this is that part of the spatio-temporal variation in \textit{CC} might be better explained when incorporating engine exhaust properties like exhaust temperature and chemical composition. Based on the aircraft type the engine types could be estimated, which in turn could yield the fuel flow (in kg/s). The total fuel combustion within a box each hour (i.e. kilograms of fuel burnt per km$^2$ per hour) could be computed by looking at the time spent for each aircraft in the respective atmospheric box and multiplying this with its fuel flow, thereby assuming usual cruising conditions.

Using the Aircraft Performance Database, a Natural Language Processing (NLP) algorithm has been built, as technical aircraft details are given in plain, unstructured text formats. The NLP algorithm is given the task to extract 1) the number of engines and 2) the engine identification number for each aircraft. The engine types, now converted into parsed strings free of special characters, are algorithmically searched for in the ICAO Aircraft Emissions Databank. Extracting the respective fuel flow from the database yields upon multiplication with the number of engines the total fuel flow of the aircraft. The flowchart of the fuel flow retrieval algorithm is shown in Figure \ref{fig:flowchartff}.

Moreover, Figure \ref{fig:flowchartff} shows at each principal edge an example for March 2015 of the total number of aircraft types that are included at that point in the analysis. The pre-processed dataset contains 130 unique aircraft types, from with 115 are found in the Aircraft Performance Database. Subsequently, seven aircraft types are lost when trying to extract the number of engines and engine type from the metadata, which is associated with the performance of the NLP algorithm. The weakest link within the system is there where the engine fuel flow should be derived from the ICAO Aircraft Emissions Databank. Here about half of the total number of aircraft engines are not recognized, which makes the model incorporating fuel flow not yet implementable. The analysis will resume with the \textit{ATD} parametrization based on Euclidean distance. Incorporation of fuel burn is left for further research.
\section{Control variable handling using ML algorithms}

To capture the variability in cirrus cover imposed by meteorological variables -- air temperature, \textit{RH} and pressure -- Machine Learning (\textit{ML}) algorithms are used. Two \textit{ML} models are built that could fit its purpose: logistic regression (\textit{LR}) and Random Forest (\textit{RF}). \textit{LR} is looked at because of its simplicity and interpretability, while \textit{RF} is looked at because of its applicability, non-parametric nature and generally good performance on unbalanced datasets. The \textit{ML} models are trained for the purpose of subsetting atmospheric boxes where \textit{CC} formation has a comparative probability of occurrence based on meteorology. This approach would expose \textit{CCC} variability caused by variations in \textit{ATD}.

The airspace is again binned into 11 layers based on data availability of ERA5 Reanalysis (centered at pressure levels of 100, 125, 150, 175, 200, 225, 250, 300, 350, 400 and 450 hPa). Gathering data for each defined $0.25^{\circ}\times 0.25^{\circ}$ grid cell over 11 pressure levels results in a dataset of around 1 million data points for Jan 2015, which is used for training and testing both models. This dataset is randomly split into a train set (80\%) and a test set (20\%), where for each data point a binary encoding (0 or 1) is used, where zeros translate into "no \textit{CC} present within this atmospheric box`` and ones translate into "(at least some) \textit{CC} present within this atmospheric box``.

\subsection{Logistic Regression Model}

The probability \textit{P} of cirrus formation based on meteorology depends particularly on whether the ambient temperature exceeds a threshold temperature, which is a function of \textit{RH} and pressure level. Looking at Figure \ref{fig:appleman} there appears to be, given \textit{RH}, a near-linear decision boundary as to where cirrus contrails may form. The \textit{LR} algorithm assumes that the logit of the probability of \textit{CC} occurrence, $\log\Big(\frac{P}{1 - P}\Big)$, is a linear function of temperature (\textit{T}), pressure (\textit{p}) and \textit{RH} (\textit{h}), see Equation \ref{eq:logit}. 
\begin{align}
    P = \frac{exp(b_0 + b_1T + b_2p + b_3h)}{1 + exp(b_0 + b_1T + b_2p + b_3h)}, \text{ or}  \label{eq:logit}
\\
    \log\Big(\frac{P}{1 - P}\Big) = b_0 + b_1T + b_2p + b_3h.\nonumber
\end{align}
It is expected that \textit{ATD} also influences the probability of cirrus occurrence. However, the location of the decision boundary is expected not to be sensitive to this boundary, as it is particularly constraint by meteorological variables. The model performance is shown in Figure \ref{fig:RFperformance}. Both test and train accuracy never exceed the constant predictor line (red dash-dotted line), which means that the algorithm performs worse than if it would be always predicted that there are no $CC$ (92\% of cases).
\subsection{Random Forest Model}

The decision tree algorithm, which forms the basis of the \textit{RF} method, works by partitioning the feature space of a dataset into subsets, where at each time the partition is performed on one variable. This is usually done in such a way that each subset is as homogeneous as possible. The \textit{RF} classification method works by combining an ensemble of decision trees into one model, whereby each sample is assigned to a certain class based on majority voting, and whereby at each feature space division a subset of features is considered. Important hyperparameters such as the maximum tree depth and the number of considered features at each split have been optimized and are set at 50 and 2, respectively. 

\begin{figure}[t]
    \centering
    \includegraphics[width=0.5\textwidth]{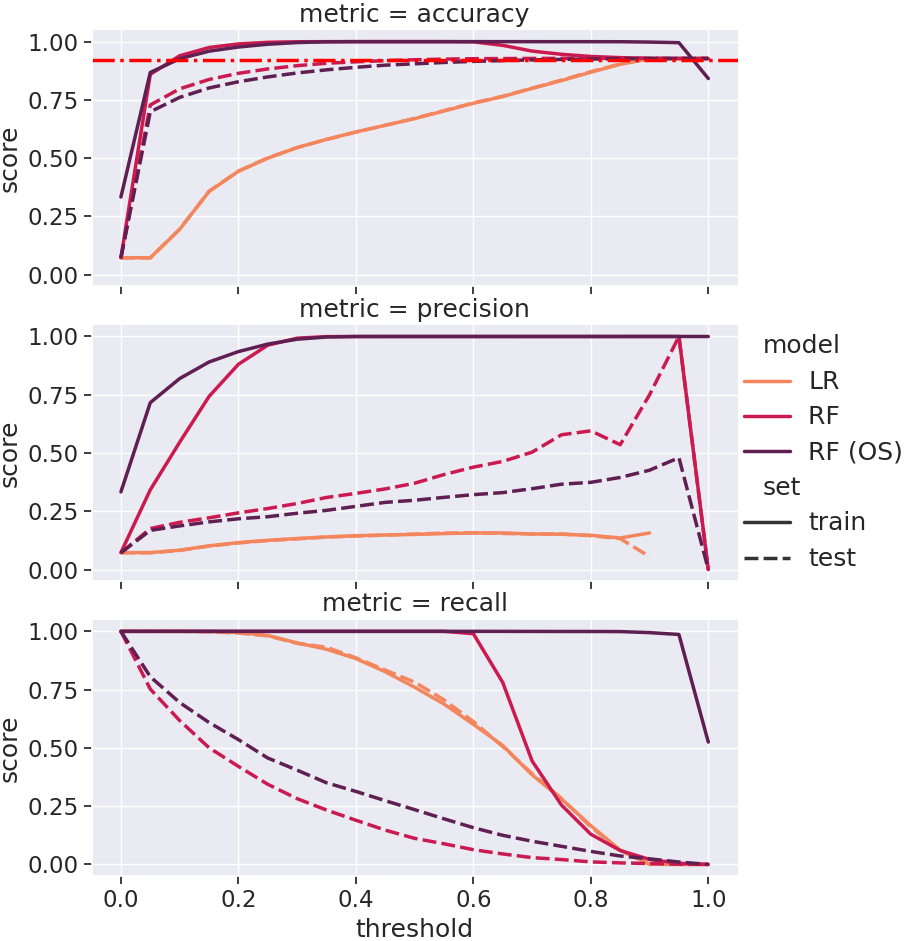}
    \caption{Performance of the \textit{LR} and \textit{RF} model, measured by the accuracy (fraction of correctly predicted cells over total), recall (fraction of correctly predicted cirrus occurrence over all actual cirrus occurrences) and precision (fraction of correctly predicted cirrus occurrence over all predicted cirrus occurrences).}
    \label{fig:RFperformance}
\end{figure}

The \textit{RF} model is run for 500 trees, where a balanced class weight is assigned to both classes in order to account for the class imbalance and the potential bias this generates in the model. In addition, the dataset is reconstructed by oversampling the minority class such that the new class balance would be two-third cirrus and one-third no cirrus. This is done in another attempt to handle the large class imbalance, potentially leading to model bias towards the majority class. Both duplicating existing instances from the minority class and generating new data belonging to the minority class (using \textit{SMOTE} \citep{chawla2002smote}) were done. Figure \ref{fig:RFperformance} shows that oversampling leads to an improvement of the model performance on the training data, but the performance on test data is low. The model performance on the train set are assessed using the Out-of-Bag (\textit{OOB}) score using a cross-validation approach. Due to the non-parametric nature of the data, the training accuracy performance is well above the constant predictor line (in red), in contrast with the \textit{LR} model. The threshold indicates the cut-off probability, i.e. the boundary which defines the separation between both classes. The test accuracy does again not exceed the accuracy of the constant predictor. It would be particularly desirable for the precision score to be higher for the test set, as this is the principal indicator for the model quality when adopted within this research. For high cut-off probabilities the precision shoots up rapidly, albeit the size of the predicted cirrus class becomes small. This is also reflected into the low recall score. The oversampling procedure leads to an improvement of the train performance, while generally decreasing the test performance. Exclusion of data on pressure levels where very little cirrus occurs does not lead to significant improvements in the model performance.

Considering the rather flawed performance of both the \textit{LR} and \textit{RF} model on unseen data, these models are not incorporated further in this research. Instead, the intuitive binning practise based on air supersaturation is applied, accompanied with meteorological statistics and a fraction related to the number of Ice Supersaturation Regions (\textit{ISSRs}) \citep{issr}. This latter quantity is computed by the dividing the number of regions where supersaturated conditions apply ($h > 100\%$) by the total number of regions.

\begin{figure*}[t]
        \centering
        \begin{subfigure}[b]{0.475\textwidth}
            \centering
            \includegraphics[width=1\textwidth]{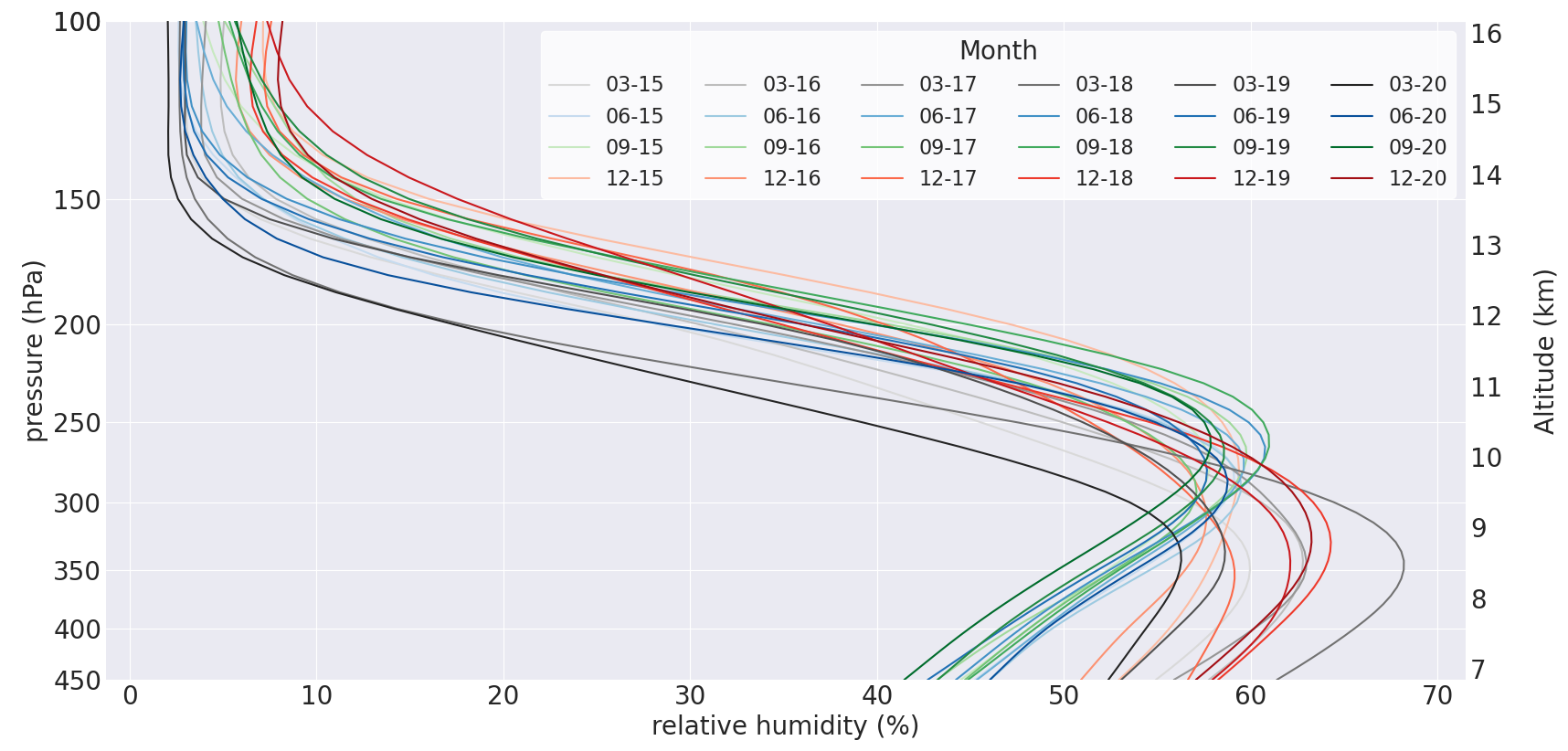}
            \caption{Vertical profile of relative humidity.}%
            \label{fig:vert_prof_RH}
        \end{subfigure}
        \hfill
        \begin{subfigure}[b]{0.475\textwidth}  
            \centering 
            \includegraphics[width=1\textwidth]{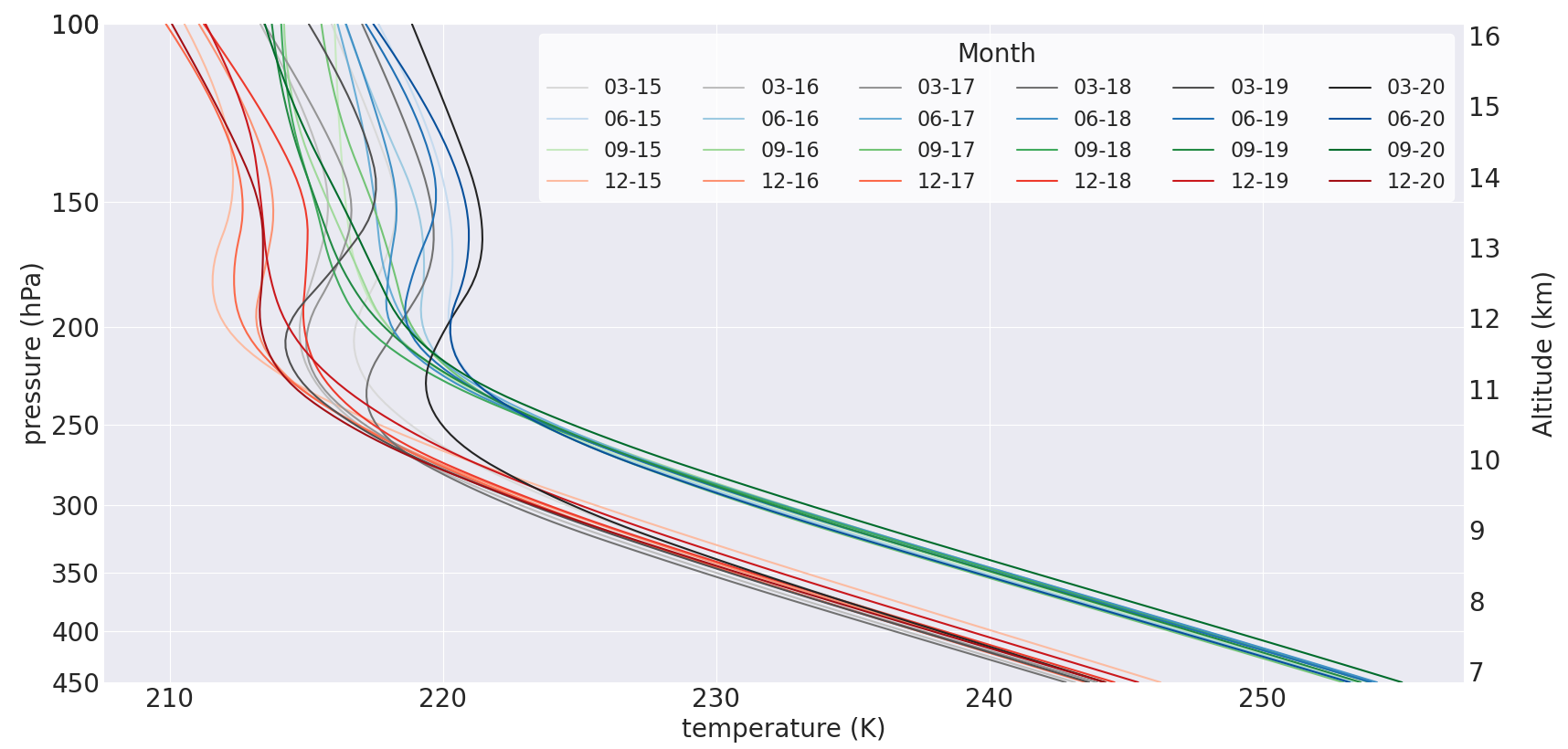}
            \caption{Vertical profile of temperature.}%
            \label{fig:vert_prof_temp}
        \end{subfigure}
        \vskip\baselineskip
        \begin{subfigure}[b]{0.475\textwidth}   
            \centering 
            \includegraphics[width=1\textwidth]{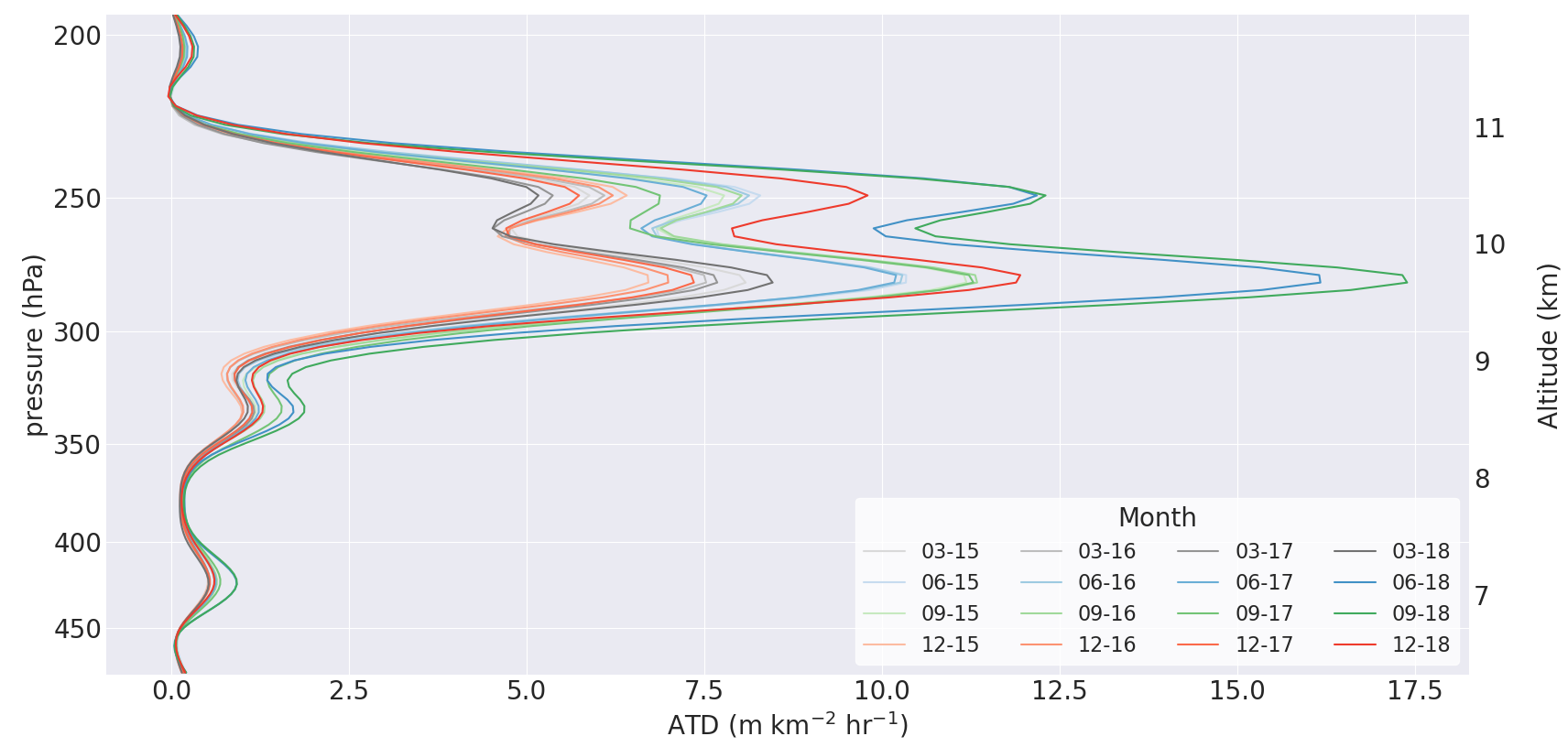}
            \caption{Vertical profile of ATD.}%
            \label{fig:vert_prof_ATD_all}
        \end{subfigure}
        \hfill
        \begin{subfigure}[b]{0.475\textwidth}   
            \centering 
            \includegraphics[width=1\textwidth]{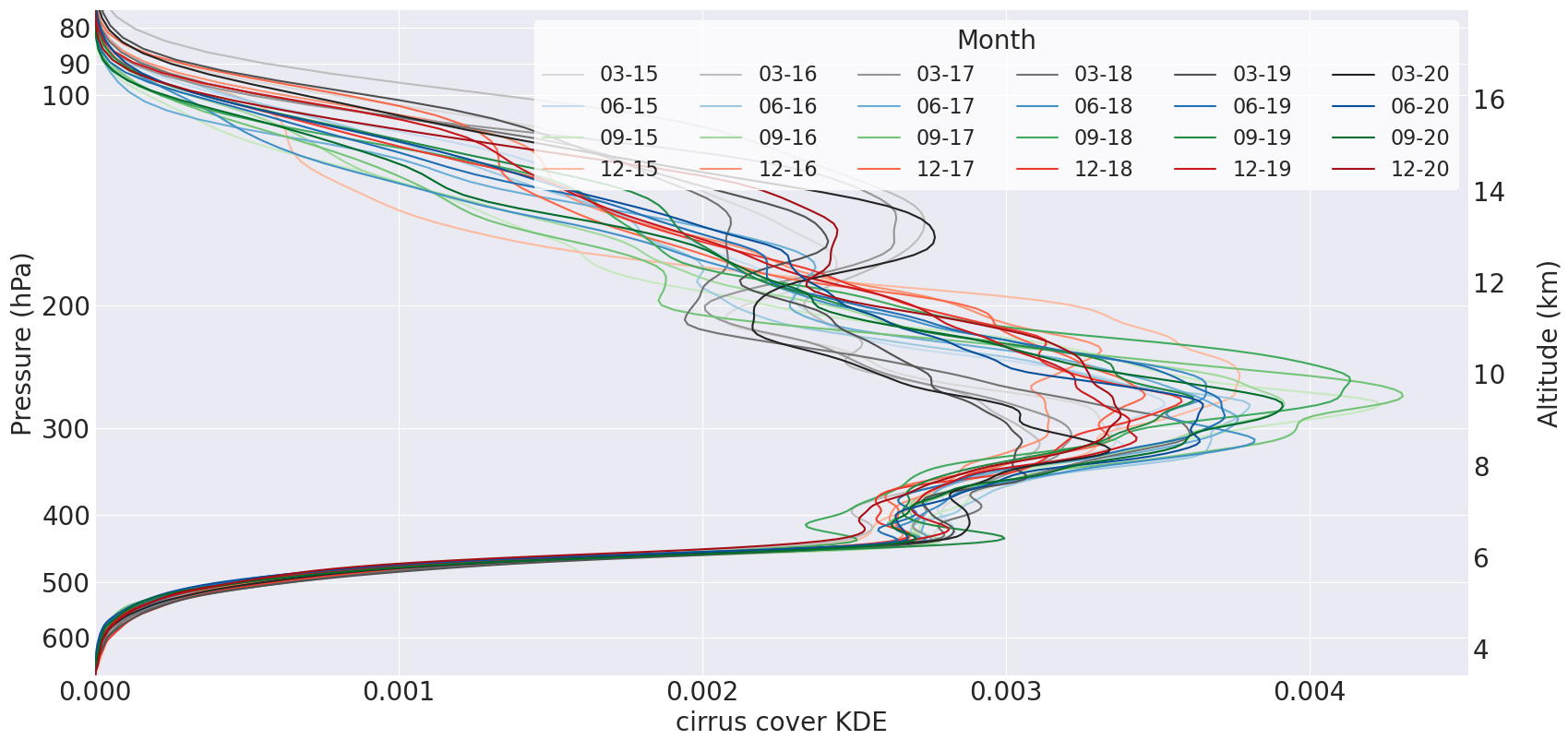}
            \caption{Vertical profile of \textit{CCC}.}%
            \label{fig:vert_prof_cirrus}
        \end{subfigure}
        \caption{Vertical \textit{RH} (a), temperature (b), \textit{ATD} (c) and \textit{CCC} (d) profiles for each month included in the analysis. Data from ERA5, EUROCONTROL and CALIPSO. The color aesthetic is used to distinguish different months. The opacity is used to indicate the year, with lower transparency for more recent years.}
        \label{fig:vert_prof_atdcirr}
\end{figure*}

\section{Correlation analysis between \textit{CCC} and flight density}

\subsection{Long-Term Time Series Analysis}

\subsubsection{Vertical Profiles}

The monthly \textit{CCC} determined by using CALIPSO data for 16 months from 2015-2020, together with the monthly \textit{ATD} and meteorological variables, are determined within each of the 11 pre-specified vertical layers. These data are spatially averaged (in the horizontal) and temporally averaged (over one month). The vertical \textit{RH} profiles from Figure \ref{fig:vert_prof_RH} have been constructed based on 11 pressure levels by cubic splines. The figure shows a peak in \textit{RH} in the upper troposphere, and a strong decline in \textit{RH} with increasing height afterwards. This is a clear tropopause signature, consistent with literature. The \textit{RH} maxima are located at lower altitudes during winter than during summer, which could be related to a seasonal vertical shift of the tropopause. The tropopause is located lower during winter than during summer, potentially impeding moisture to rise further upwards. Also, the maxima are higher in magnitude during winter. March 2020 was an exceptional year as the \textit{RH} maximum was even lower than usual during the summer months, while the temperature above 250 hPa was also anomalously high.

The vertical temperature profiles from Figure \ref{fig:vert_prof_temp} are piece-wise linear regression graphs based on temperature data attained on 11 pressure levels. Piece-wise linear regression is chosen as temperature is expected to vary at a constant lapse rate with height. The temperature decreases with height at a seemingly constant lapse rate in the upper troposphere, consistent with literature, and the upper tropospheric mean temperature during summer is clearly offset compared to winter with about 10K. In the tropopause and lower stratosphere the temperature profiles are ostensibly not strongly affected by the seasonal cycle. However, a discrepancy with theory arises at lower stratospheric temperatures, where temperature first increases with height as expected and subsequently drops again. This points to some bias in the data within the lower stratosphere.

In Figure \ref{fig:vert_prof_ATD_all} the vertical \textit{ATD} profiles are shown. Clearly most air traffic is flying between 9.5 and 10.5 km altitude, indifferent of the month. Air traffic clearly increases in magnitude over the course of four years, consistent with Figure \ref{fig:air_traffic_trend}.

The vertical \textit{CCC} distribution derived from CALIPSO is shown in Figure \ref{fig:vert_prof_cirrus}. Consistent with literature the vast majority of \textit{CC} occur between 6 and 14 km height. The kernel density estimate emphasizes that the highest proportion of cirrus is occurring around 10 km altitude, where the vertical \textit{ATD} maxima (see Figure \ref{fig:vert_prof_cirrus}) coincide with the \textit{RH} maxima. This is most evident for the summer months (June and September). During winter months the vertical cirrus distribution between 6 and 14 km is more uniform, possibly because of tropopause propagation downwards and relatively lower \textit{ATD} at those flight levels where the \textit{RH} attains its maximum. The validity of this proposition should be further researched.

\subsubsection{Long-term \textit{CC}}

\begin{figure}[t]
    \centering
    \includegraphics[width=0.5\textwidth]{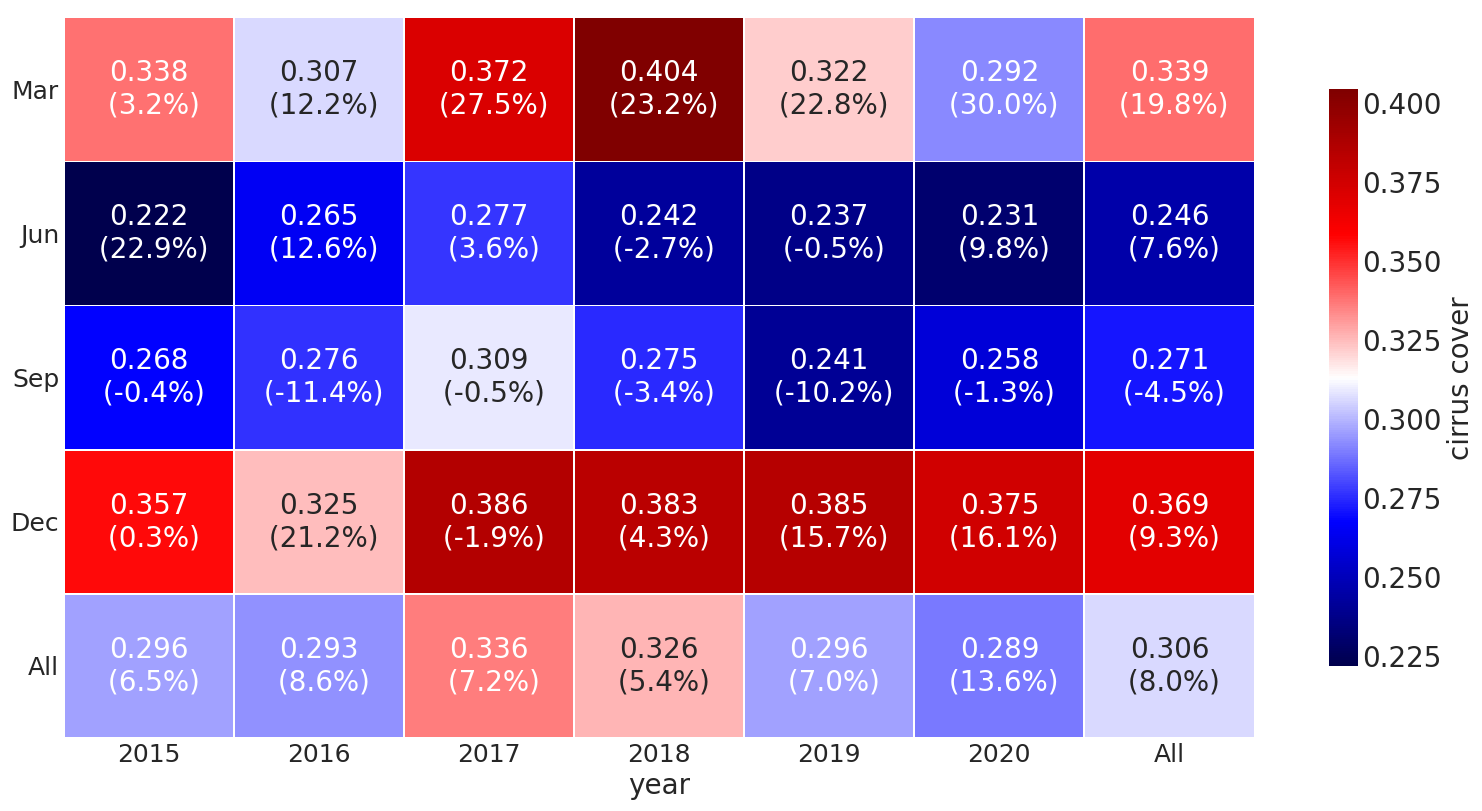}
    \caption{Heat map of \textit{CCC} for the CALIPSO CALIOP data. Time series runs from 2015 till 2020. The percentages show the difference between daylight and night \textit{CCC}. The cover during daylight is taken as reference.}
    \label{fig:CALIPSO_timeseries_inclpercentage}
\end{figure}

Figure \ref{fig:CALIPSO_timeseries_inclpercentage} shows a heat map of monthly average \textit{CCC} for all months included in the analysis. Months are shown on the vertical, years on the horizontal. There is a consistent pattern that \textit{CCC} is higher during the winter months (March and December) than during summer months, with a low anomaly in March 2020 and December 2016, and a high anomaly for September 2017. This general pattern could be highly associated with meteorological variables, particularly \textit{RH} and temperature, as shown in Figures \ref{fig:vert_prof_RH} and \ref{fig:vert_prof_temp}. Over the evaluated time span Figure \ref{fig:CALIPSO_timeseries_inclpercentage} provides no indication of an upward trend in \textit{CCC} due to air traffic increase. Mean \textit{CCC} was lowest over 2020 when the COVID-19 outbreak took place, even though the difference with e.g. 2016 is small. A low \textit{RH} maximum in combination with high temperatures above 250 hPa and a strong decline in \textit{ATD} (see Figure \ref{fig:air_traffic_trend}) could explain the clear drop in \textit{CCC} seen in Figure \ref{fig:CALIPSO_timeseries_inclpercentage}.

Figure \ref{fig:CALIPSO_timeseries_inclpercentage} shows the day-night difference in $CCC$ percentage-wise, taking the day cover as reference (i.e. positive percentage means higher cover during the nights than during the day). It is evident from this figure that \textit{CCC} is generally higher during the night than during the day (respective means of 0.320 and 0.295 or 8\% difference). For all years in September the opposite is the case. Apart from the possible linkage with synoptic-scale meteorology, the role of air traffic cannot be precluded based on this analysis, as \textit{AIC} should predominantly occur during daytime. In fact, out of the four considered months, \textit{ATD} is highest in September (Figure \ref{fig:air_traffic_trend}) whilst the geopotential height of the \textit{RH} maxima in those months (Figure \ref{fig:vert_prof_RH}) closely coincides with the predominant flight level (Figure \ref{fig:vert_prof_ATD_all}).

\subsubsection{Long-term \textit{ATD}}

Figure \ref{fig:air_traffic_trend} shows the time series analysis on the monthly number of flights over the \textit{ROI}. The number of flights in 2019 and 2020, not shown in this figure, were retrieved based on reported geolocations and on airport combinations as explained in Section \ref{sec:atspatio-temp}. This retrieval algorithm tends to underestimate the \textit{ATD}, relating to the data quality and completeness of those datasets. A later publication of the full European air traffic data files for March 2019 on the EUROCONTROL website made it possible to evaluate the number of flights that had been taken place during this month, assuming this source to be most reliable. From here it resulted that the estimated number of flights for March 2019 using the inferior dataset is 89\% of the true value, i.e. an under-estimation of the number of flights of 11\%. 

Clearly there are more flights taking place during the summer months than during winter. This seeming anti-correlation between \textit{ATD} and \textit{CCC} can be explained by the dominant effect of meteorology, overshadowing the signatures left by air traffic. This is partially confirmed by Figures \ref{fig:vert_prof_RH} and \ref{fig:vert_prof_temp} which show consistently lower upper tropospheric temperatures (between approximately 450 hPa and 200 hPa) in winter, and in December also higher \textit{RH}, both favorable for cirrus formation. Also, air traffic shows a consistent increase around 4\% per year, with a sharp decline around 51\% in 2020 relative to 2019. This variability cannot be seen back in the mean \textit{CCC} over this time period.

\subsubsection{Meteorology-based Data Subsetting}

\begin{figure}[t]
    \centering
    \includegraphics[width=0.5\textwidth]{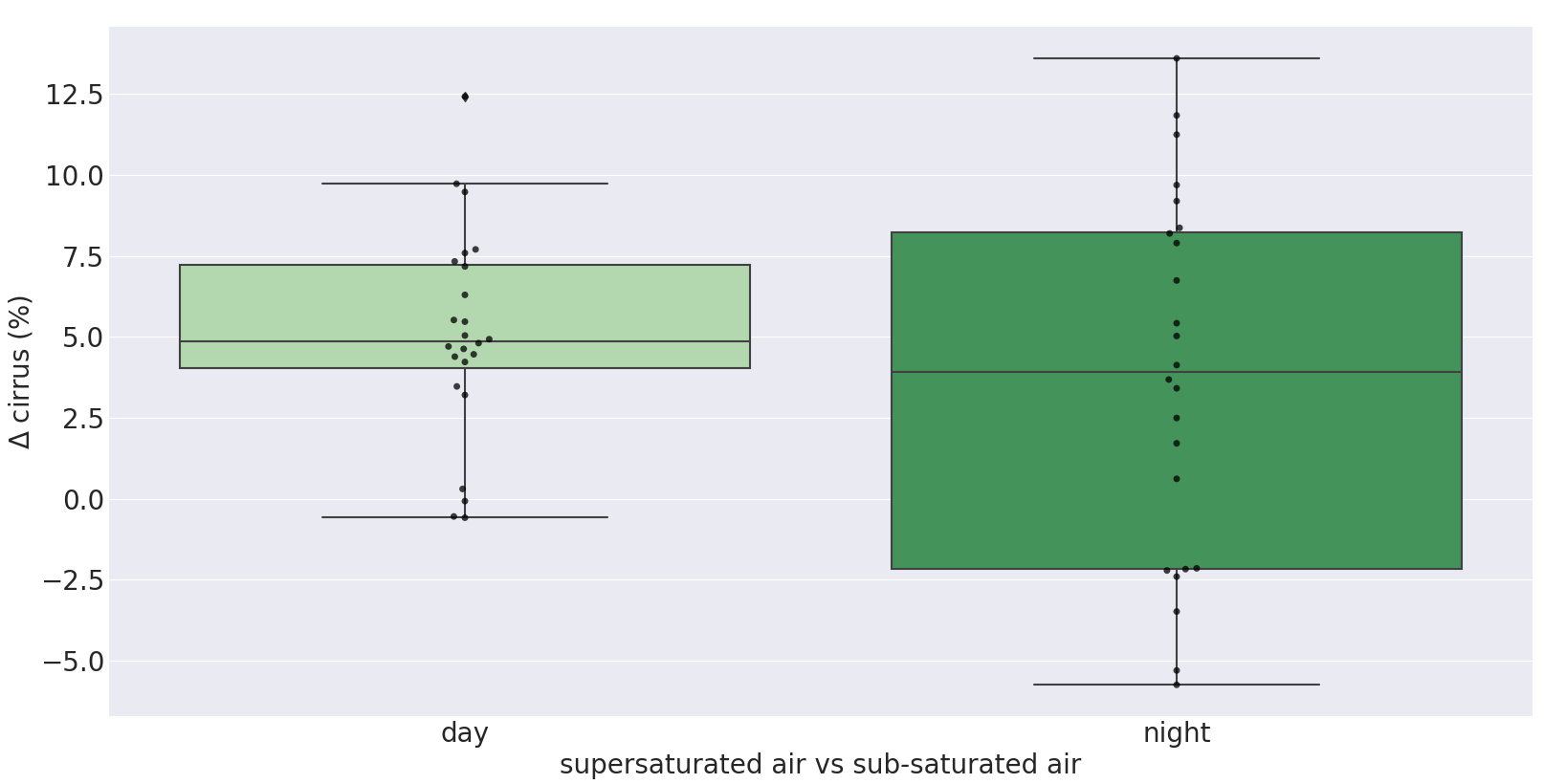}
    \caption{Side-by-side boxplot of difference in \textit{CCC} between sub-saturated and supersaturated air, subset on day and night cover.}
    \label{fig:boxplot_cirrus}
\end{figure}

To eliminate part of the variability induced by meteorological conditions, a raw data grouping is done based on air saturation level. A comparison of monthly \textit{CCC} between supersaturated and sub-saturated air is shown in Figure \ref{fig:boxplot_cirrus}, where the data has been split based on daytime. I.e. the left boxplot shows the percentage difference in \textit{CCC} between sub-saturated air and supersaturated air during daytime, while the right boxplot shows the difference during nighttime. Clearly there is a higher spread in differences over night than during the day, and the difference of the means is statistically significant at a 99\% confidence level with respective means of 3.7\% and 5.1\%. Diurnal temperature and \textit{RH} cycles are weak in the upper troposphere and lower stratosphere and do not explain the observed day-night differences. A plausible explanation for the observed day-night difference is an influence of air traffic flying during daytime in supersaturated air, increasing the difference in \textit{CCC} between supersaturated and sub-saturated regions, compared to the night where air traffic is mainly absent.

To further elaborate on the question what is causing the observed fluctuations in monthly mean \textit{CCC}, the fraction is \textit{ISSR} regions is calculated for each month. Also the \textit{ISSR} fractions show a significant difference between summer and winter. The relatively low (high) mean \textit{CCC} in March 2020 (September 2016) coincides with a relatively low (high) \textit{ISSR} fraction. The same applies to December 2016. The R$^2$ between \textit{ISSR} and CALIPSO monthly \textit{CCC} is 0.857, therefore explaining about 86\% of the observed \textit{CCC} variance.

\subsection{\textit{CCC} Analysis on Shorter Timescales}

\subsubsection{Short-Term Time Series Analysis}

\begin{figure}[t]
\centering
\subcaptionbox{Time series of ATD and \textit{CCC}.\label{fig:timeseriesATD_cirrus_Mar15}}{%
  \includegraphics[width=0.5\textwidth]{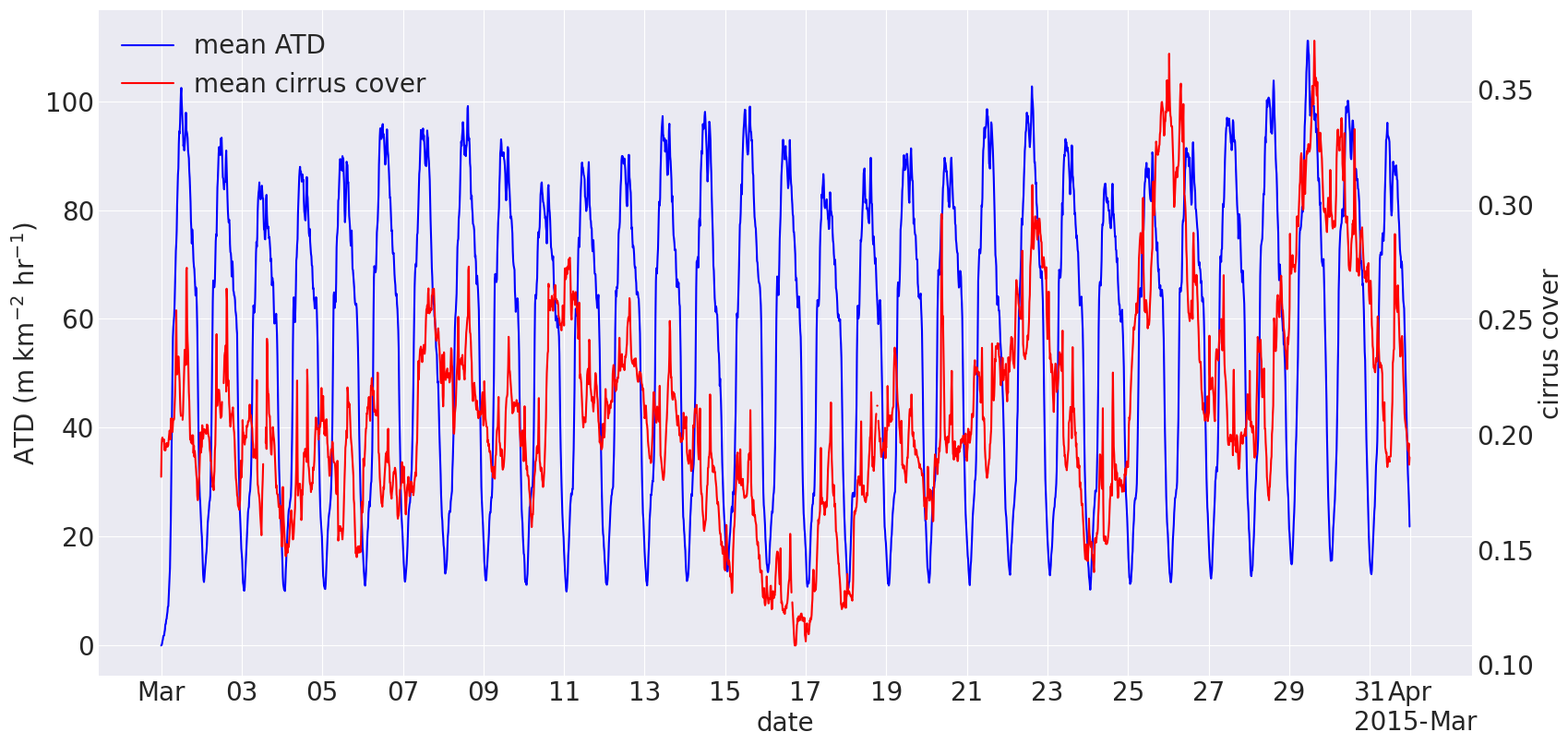}%
  }\par\medskip
 \subcaptionbox{ISSR plotted as a percentage of all regions for March 2015. In the background the \textit{CCC} trend is plotted.\label{fig:ISSR_time_series}}{%
  \includegraphics[width=0.5\textwidth]{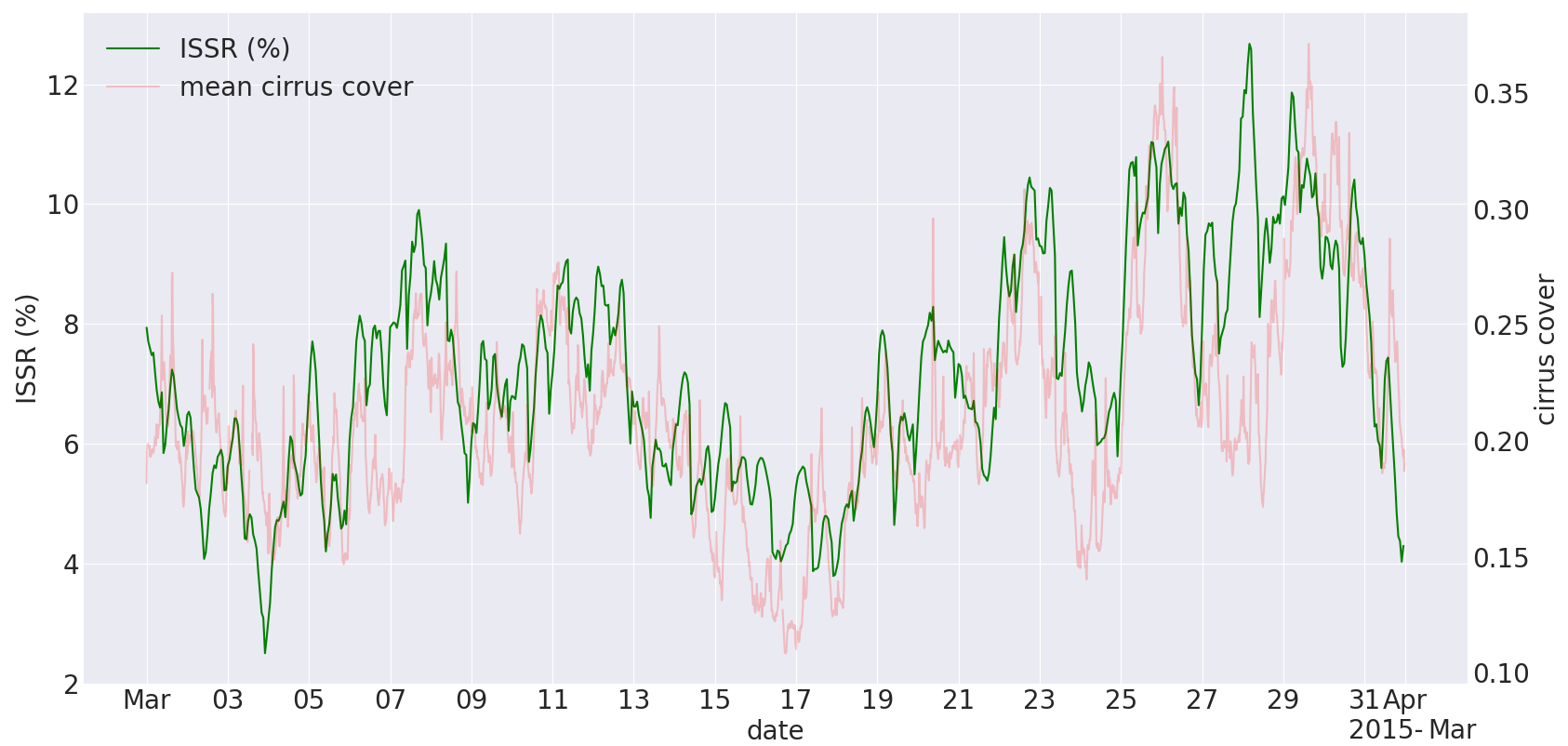}
  }
\caption{ATD vs \textit{CCC} for March 2015.}
\label{fig:atdvscirrusmar2015}
\end{figure}

Spatially averaged flight data from March 2015, evaluated every 15 mins have been plotted in Figure \ref{fig:timeseriesATD_cirrus_Mar15} against time. The daily air traffic cycle is clearly visible with peaks during daytime around 100 m km$^{-2}$ hr$^{-1}$, dropping to around 10 m km$^{-2}$ hr$^{-1}$ at night. A weekly cycle is perceived which relates to more scheduled flights during the weekends, with the highest peaks attained on Sundays and the lowest on Tuesdays.

The spatially averaged \textit{CCC} shown in red, retrieved from the METEOSAT data product, results as a much noisier signal with some apparent synoptic variability over time scales of a couple of days to a couple of weeks. This synoptic variability cannot be explained by air traffic, and appears to be dictated by synoptic meteorological conditions. This is illustrated in Figure \ref{fig:ISSR_time_series}, where the fraction of \textit{ISSR} is plotted against time. On the background the mean \textit{CCC} is shown. The mean \textit{CCC} strongly follows the \textit{ISSR} pattern, and the latter explains over half (R$^2 = 0.52$) of the \textit{CCC} variability. 

\subsubsection{Short-term Correlation between \textit{ATD} and \textit{CCC}}

The datasets are binned based on \textit{ATD} using the 5-point Likert scale - "\textit{very low ATD}", "\textit{low ATD}", "\textit{moderate ATD}", "\textit{high ATD}" and "\textit{very high ATD}". The Jenks optimization method is used to achieve this. This method seeks for natural splits in uni-variate data whereby it tends to minimize the intracluster variability and maximize the intercluster variability. The method is not highly scalable. Therefore, the algorithm is run on separate time windows during the day. The class boundaries obtained during each time window are averaged to obtain the final five clusters that are applied to the entire dataset. As with K-means clustering, a drawback of the Jenks optimization method is that the number of clusters should be given to the algorithm, a choice that in most cases might be arbitrary.

\begin{figure}[t]
    \centering
    \includegraphics[width=0.5\textwidth]{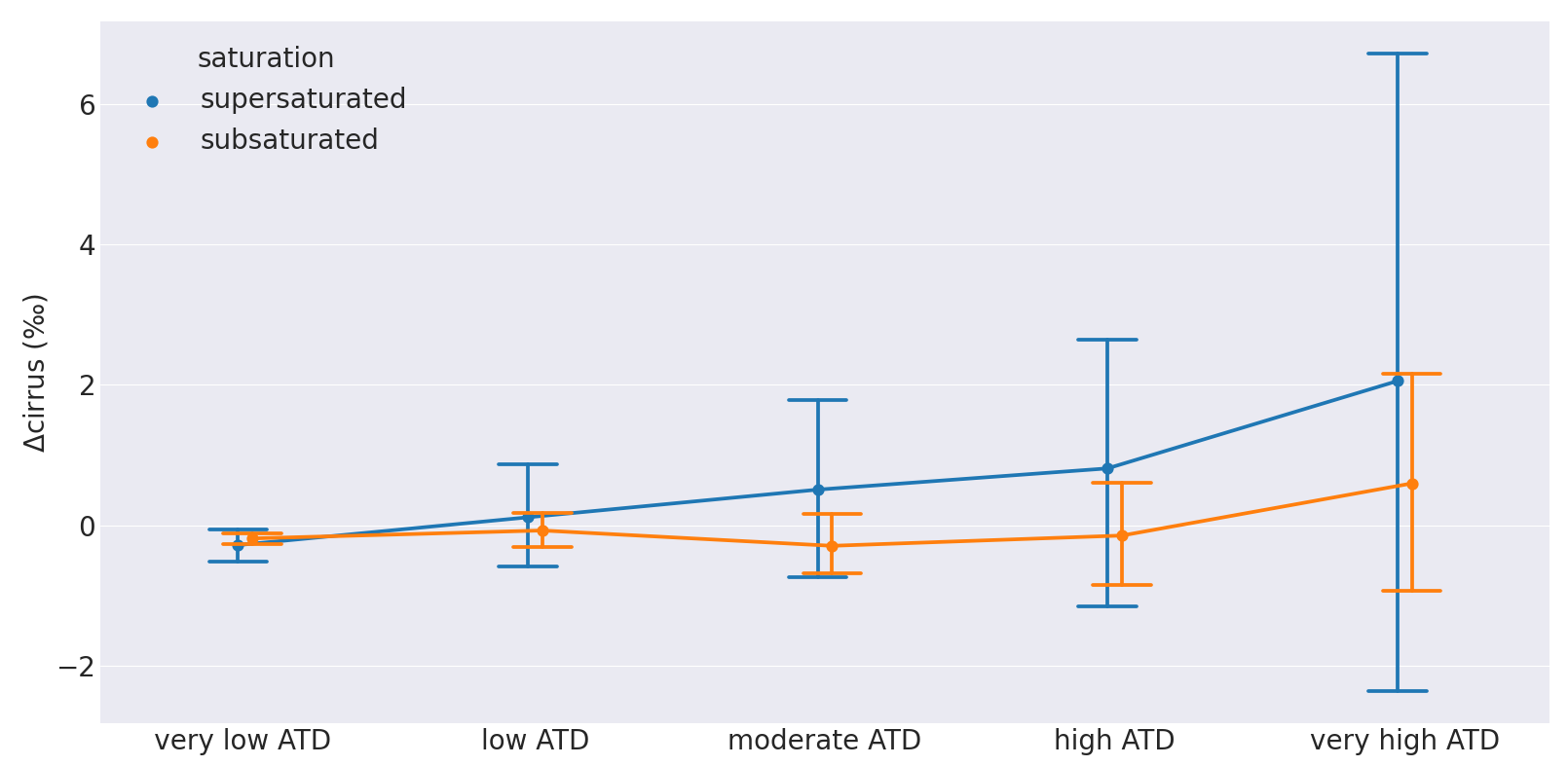}
    \caption{The mean CCC increment in permilles within the five different Air Traffic Density bins. Different plots are shown for sub-saturated ($RH < 100\%$) and supersaturated ($RH > 100\%$) air. Error bars show the 95\% confidence intervals based on a bootstrapping technique with $n = 1000$.}
    \label{fig:trend_saturations}
\end{figure}

Subsequently each class is subset into supersaturated and sub-saturated regions, and the mean percentage point changes in \textit{CCC} over one sampling period (15 mins) for each subset is determined. This is shown in Figure \ref{fig:trend_saturations}. The error bars indicate the 95\% confidence intervals for each class mean based on the bootstrapping technique, with 1,000 abstractions.

Both trends from Figure \ref{fig:trend_saturations} are tested for significance using a t-test with the significance level set at 0.05. Strong evidence is found for a significant positive correlation between \textit{ATD} and \textit{CCC} change under supersaturated conditions, with a slope of 11.2 percentage points per km km$^{-2}$ hr$^{-1}$ and a $p$-value of $6\cdot 10^{-5}$. In sub-saturated air however, no highly convincing support for a trend was found with a $p$-value of 0.06.

\section{Conclusion}

This study investigated the \textit{CCC} properties in the European airspace in conjunction with air traffic, by combining multiple data sources into a big data approach. Both the strengths and challenges of combining various data sources are shown. We also showed that air traffic is subject to a consistent rise in volume of about 4\% a year within the investigated airspace, leaving the question whether this gets reflected into \textit{CCC}, the latter having relevance for Earth's radiative budget. No clear evidence was found using CALIPSO that mean \textit{CCC} changes along with air traffic over the period 2015-2020.
Some signatures of air traffic were found when subsetting atmospheric regions on saturation level and making inter-comparisons. There was found a statistically significant difference between the mean difference in cirrus cover over supersaturated and sub-saturated regions during daytime and nighttime.

Looking at changes in \textit{CCC} over time scales of tens of minutes resulted in a significant correlation with \textit{ATD} in supersaturated regions, a correlation that was not found as evidently in sub-saturated regions. From this and the long-term analysis it is concluded that lifetime of contrails might play a key role here, as correlations between air traffic and cirrus cover appear mostly on shorter timescales.

Further elaboration on the research questions require higher data quality, 1) on the side of air traffic data for the years 2019 and 2020, and 2) on the linkage between aircraft metadata and engine fuel flow, which would be beneficial for the performance of the \textit{ML} algorithms. Parametrizing air traffic with incorporation of exhaust properties will increase the accuracy, as engine properties are under ongoing development. 

\bibliographystyle{ACM-Reference-Format}
\bibliography{bibliography.bib}

\end{document}